\documentclass[12pt]{article}
\usepackage[dvips]{graphicx}
\usepackage{amssymb,amsmath,fullpage}
\usepackage[latin1]{inputenc}
\begin{document}
%
%

\noindent SOME RESULTS FOR BETA FRÉCHET DISTRIBUTION \vskip 3mm

\vskip 5mm
\noindent Wagner Barreto-Souza\footnote{ E-mail: wagnerbs85@hotmail.com}$^*$, Gauss M. Cordeiro\footnote{Corresponding author. E-mail: gausscordeiro@uol.com.br}$^\dag$ and Alexandre B. Simas\footnote{ E-mail: alesimas@impa.br}$^+$\\

\noindent $^*$ Departamento de Estat\' \i stica,\\
Universidade Federal de Pernambuco,\\
Cidade Universitária, 50740-540 -- Recife, PE, Brazil\\

\noindent $^\dag$  Departamento de Estat\'{\i}stica e Inform\'atica,\\
Universidade Federal Rural de Pernambuco,\\
Rua Dom Manoel de Medeiros s/n, 50171-900 -- Recife, PE, Brazil\\

\noindent $^+$ Associa\c c\~ao Instituto Nacional de Matem\'atica Pura e Aplicada,\\
Estrada Dona Castorina 110, Jardim Bot\^anico,\\
22460-320 -- Rio de Janeiro, RJ, Brazil\\

\noindent ABSTRACT\\
Nadarajah and Gupta (2004) introduced the beta Fréchet (BF)
distribution, which is a ge\-ne\-ra\-lization of the exponentiated
Fréchet (EF) and Fréchet distributions, and obtained the
pro\-ba\-bility density and cumulative distribution functions.
However, they do not investigated its moments and the order
statistics. In this paper the BF density function and the density
function of the order statistics are expressed as linear
combinations of Fréchet density functions. This is important to
obtain some mathematical properties of the BF distribution in terms
of the corresponding properties of the Fréchet distribution. We
derive explicit expansions for the ordinary moments and L-moments
and obtain the order statistics and their moments. We also discuss
maximum likelihood estimation and calculate the information matrix
which was not known. The information matrix is easily numerically
determined. Two applications to real data sets are given to
illustrate the potentiality of this distribution.

 \vskip 4mm

\noindent 1.   INTRODUCTION \\

The Fréchet distribution has applications ranging from accelerated
life testing through to earthquakes, floods, horse racing, rainfall,
queues in supermarkets, sea currents, wind speeds and track race
records. Kotz and Nadarajah (2000) give some applications in their
book. In this paper, we discuss the BF distribution which stems from
the following idea. Eugene et al. (2002) defined the beta $G$
distribution from a quite arbitrary cumulative distribution function
(cdf) $G(x)$ by
\begin{equation}\label{distfun}
F(x)=\frac{1}{B(a,b)}\int_0^{G(x)}\omega^{a-1}(1-\omega)^{b-1}d\omega,
\end{equation}
where $a>0$ and $b>0$ are two additional parameters whose role is to introduce
skewness and to vary tail weight and $B(a,b)=\int_0^1 \omega^{a-1}(1-\omega)^{b-1}d\omega$
is the beta function. The class of distributions (\ref{distfun}) has an
increased attention after the works by Eugene et al. (2002) and Jones (2004).
Application of $X=G^{-1}(V)$ to the random variable $V$
following a beta distribution with parameters $a$ and $b$, $V\sim B(a,b)$
say, yields $X$ with cdf (\ref{distfun}).

Eugene et al. (2002) defined the beta normal (BN) distribution by
taking $G(x)$ to be the cdf of the normal distribution and derived
some of its first moments. General expressions for the moments of
the BN distribution were derived (Gupta and Nadarajah, 2004).
Nadarajah and Kotz (2004) also introduced the beta Gumbel (BG)
distribution by taking $G(x)$ to be the cdf of the Gumbel
distribution and provided closed-form expressions for the moments,
the asymptotic distribution of the extreme order statistics and
discussed the maximum likelihood estimation procedure. Nadarajah and
Gupta (2004) introduced the BF distribution by taking $G(x)$ to be
the Fréchet distribution, derived the analytical shapes of the
probability density function (pdf) and the hazard rate function and
calculated the asymptotic distribution of the extreme order
statistics. However, they do not investigate expressions for its
moments and the information matrix which we do in this paper. Also,
Nadarajah and Kotz (2005) worked with the beta exponential (BE)
distribution and obtained the moment generating function, the first
four cumulants, the asymptotic distribution of the extreme order
statistics and discussed the maximum likelihood estimation. We can
write (\ref{distfun}) as
\begin{equation}\label{cdf_*A}
F(x)= I_{G(x)}(a,b),
\end{equation}
where $I_{y}(a,b) = B(a,b)^{-1}\int_{0}^{y}w^{a-1}(1-w)^{b-1}dw$ denotes
the incomplete beta function ratio, i.e., the cdf of the beta distribution with
parameters $a$ and $b$. For general $a$ and $b$, we can express (\ref{cdf_*A}) in
terms of the well-known hypergeometric function defined by
$$_2F_{1}(\alpha,\beta,\gamma;x)= \sum_{i = 0}^{\infty}\frac{(\alpha)_{i}(\beta)_{i}}{(\gamma)_{i}i!}\,x^{i},$$
where $(\alpha)_{i}=\alpha (\alpha+1)\ldots(\alpha+i-1)$ denotes the ascending factorial.
We obtain
$$F(x)=\frac{G(x)^a}{a\,B(a,b)}\,_2F_{1}(a,1-b,a+1;G(x)).$$
The properties of the cdf $F(x)$ for any beta $G$ distribution
defined from a parent $G(x)$ in (\ref{distfun}), could, in
principle, follow from the properties of the hypergeometric function
which are well established in the literature; see, for example,
Section 9.1 of Gradshteyn and Ryzhik (2000).

The probability density function (pdf) corresponding to
(\ref{distfun}) can be written in the form
\begin{equation}\label{beta1_f}
f(x)=\frac{1}{B(a,b)}G(x)^{a-1}\{1-G(x)\}^{b-1} g(x),
\end{equation}
where $g(x)=dG(x)/dx$ is the pdf of the parent distribution. The pdf $f(x)$ will be most
tractable when the functions $G(x)$ and $g(x)$ have simple analytic expressions
as is the case of the Fréchet distribution. Except for some special choices for
$G(x)$ in (\ref{distfun}), it would appear that the pdf $f(x)$
in (\ref{beta1_f}) will be difficult to deal with.

The cdf and pdf of the Fréchet distribution are, respectively,
\begin{equation}\label{cdffrechet}
G_{\sigma,\lambda}(x)= e^{-(\frac{\sigma}{x})^{\lambda}},\quad x>0,
\end{equation}
and
$$g_{\sigma,\lambda}(x)=\lambda\sigma^\lambda x^{-(\lambda+1)}e^{-(\frac{\sigma}{x})^\lambda},\quad x>0,$$
where $\sigma>0$ is the scale parameter and $\lambda>0$ is the shape parameter.
The $r$th moment of the Fréchet distribution for $r<\lambda$ is
$\mu_r^{\prime}=\sigma^r\Gamma(1-r/\lambda)$, and then the first four cumulants if $\lambda>4$ are
$$\kappa_1 = \sigma g_1,\quad \kappa_2=\sigma^2(g_2-g_1^2),$$
$$\kappa_3 =\frac{g_3-3g_1g_2+2g_1^2}{(g_2-g_1^2)^{3/2}},\quad \kappa_4=\frac{g_4-4g_1g_3+6g_1^2g_2-3g_1^4}{(g_2-g_1^2)^2},$$
where $g_k=\Gamma(1-k/\lambda)$ for $k=1,\ldots,4$.

\begin{figure}[h!]
\centering
\includegraphics[width=0.5\textwidth]{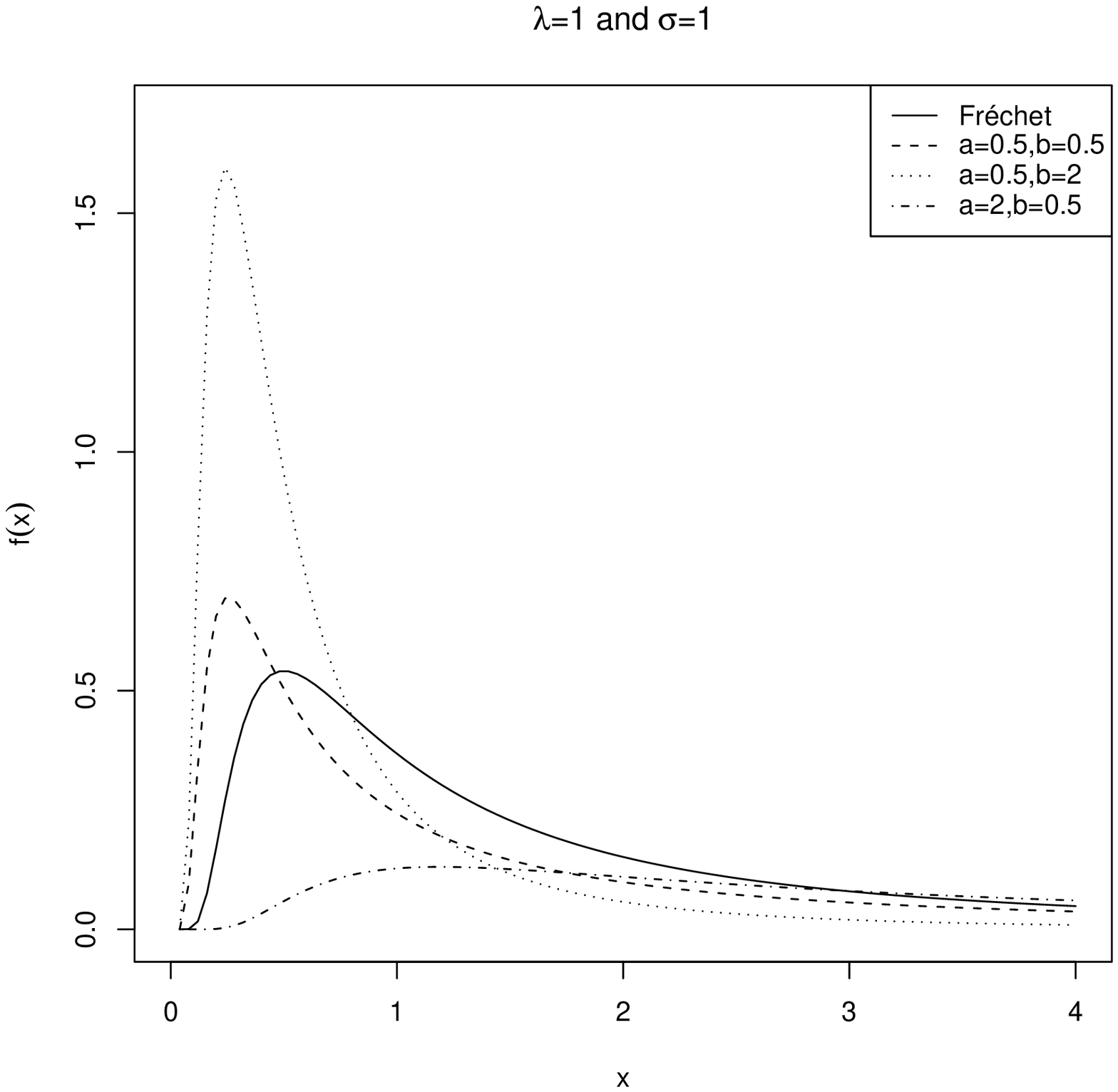}\includegraphics[width=0.5\textwidth]{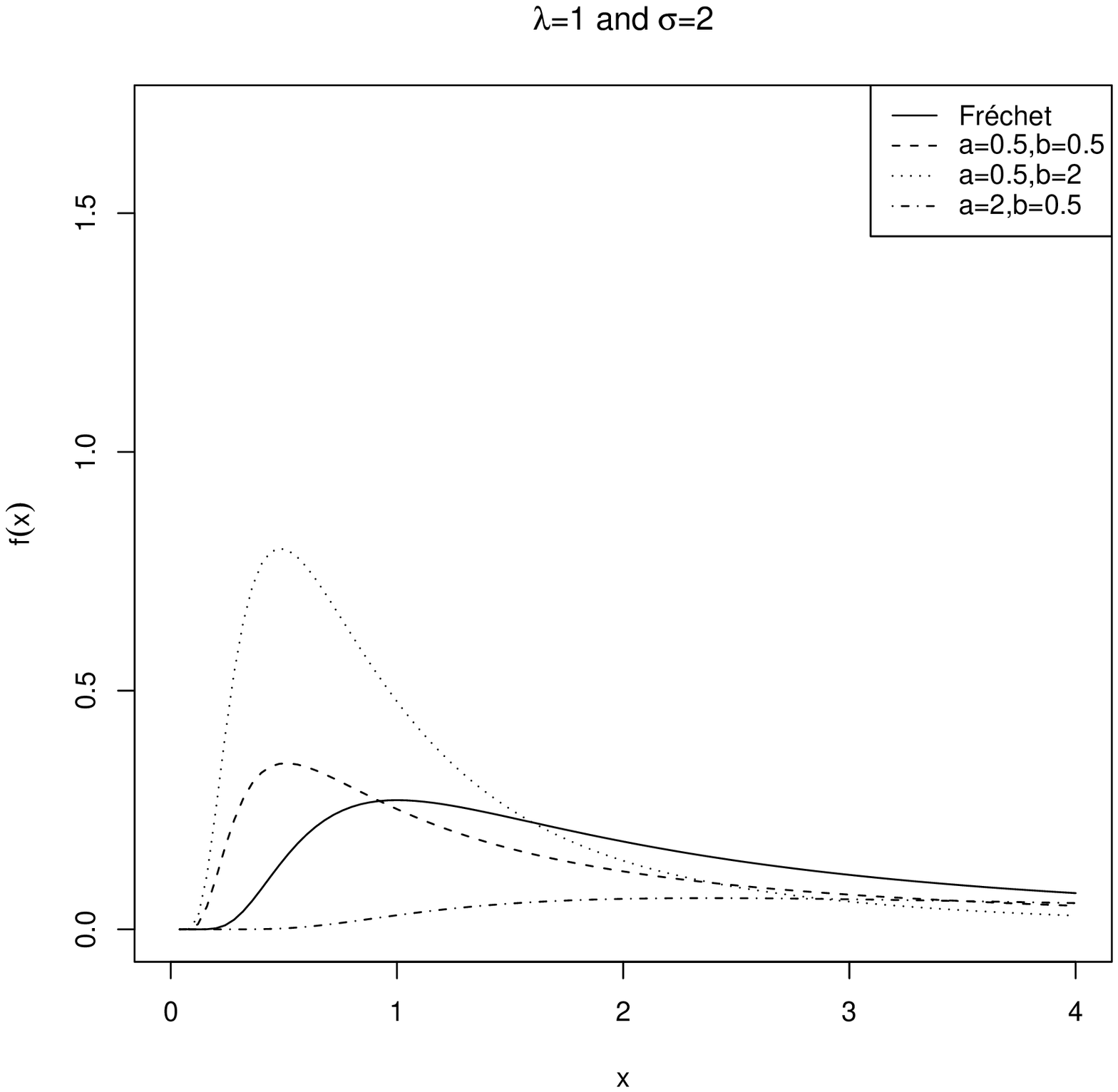}
\includegraphics[width=0.5\textwidth]{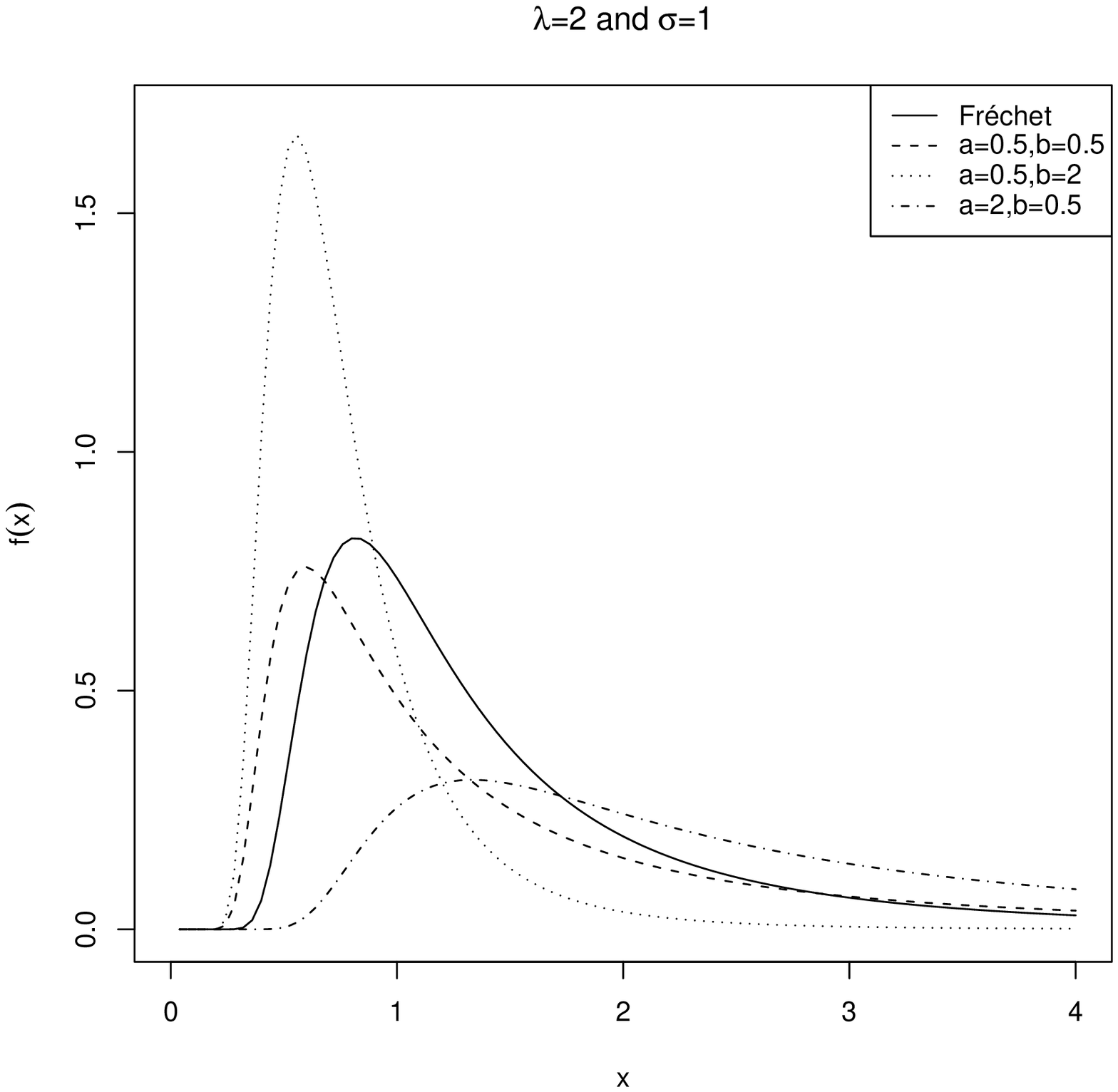}\includegraphics[width=0.5\textwidth]{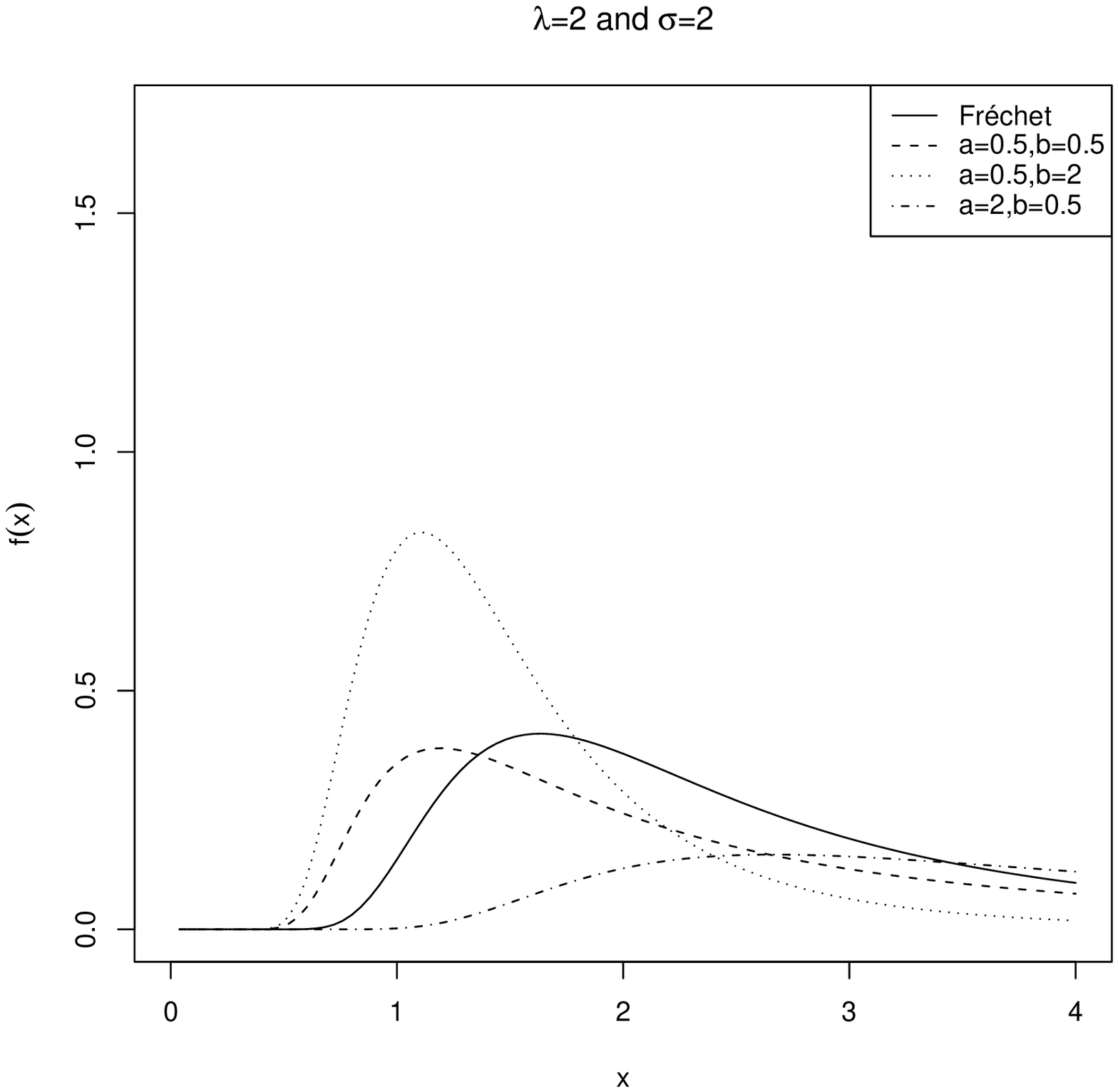}
\caption{Pdf of the BF distribution for selected parameter values.}
\label{figpdf}
\end{figure}

\begin{figure}[h!]
\centering
\includegraphics[width=0.5\textwidth]{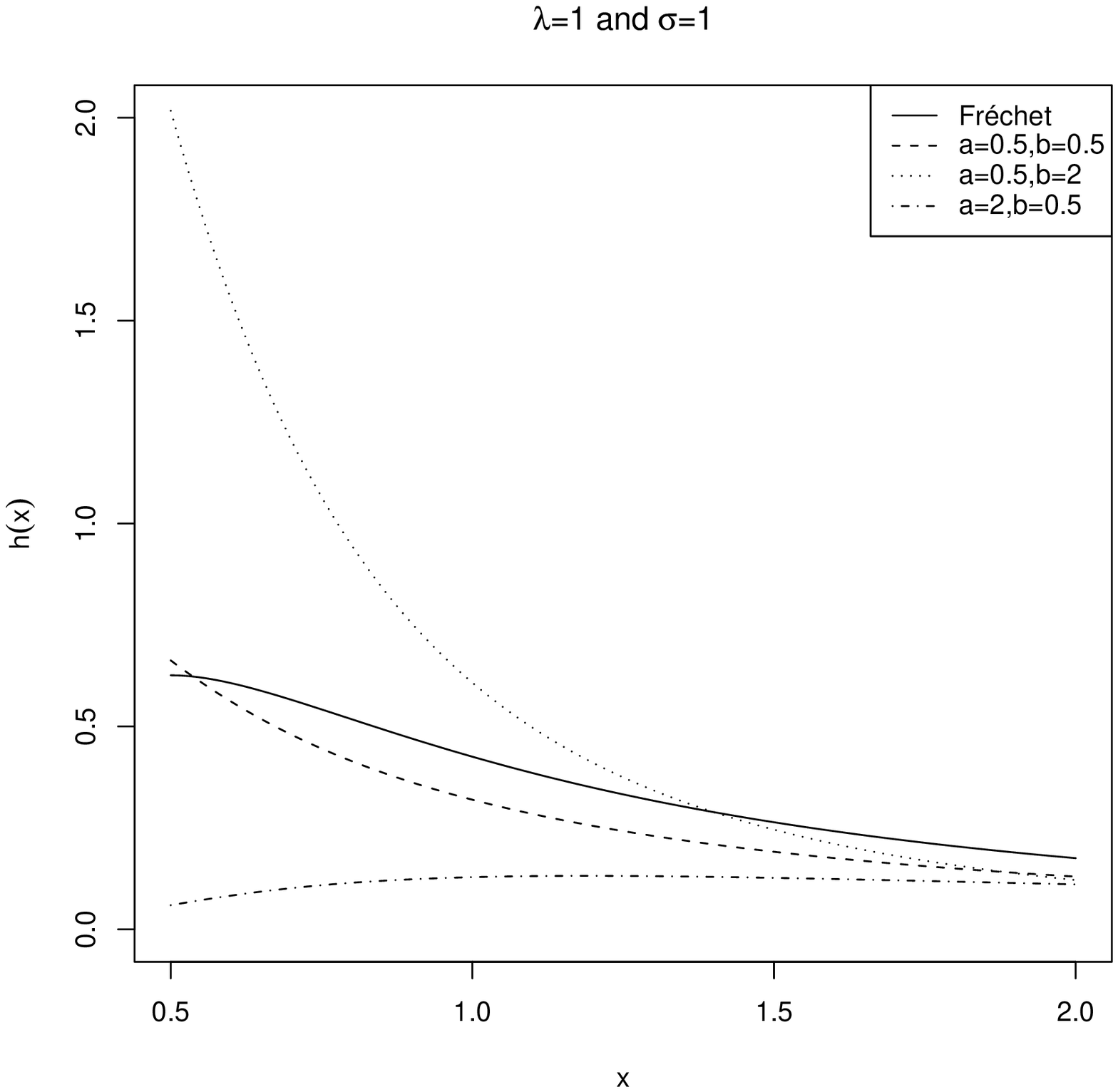}\includegraphics[width=0.5\textwidth]{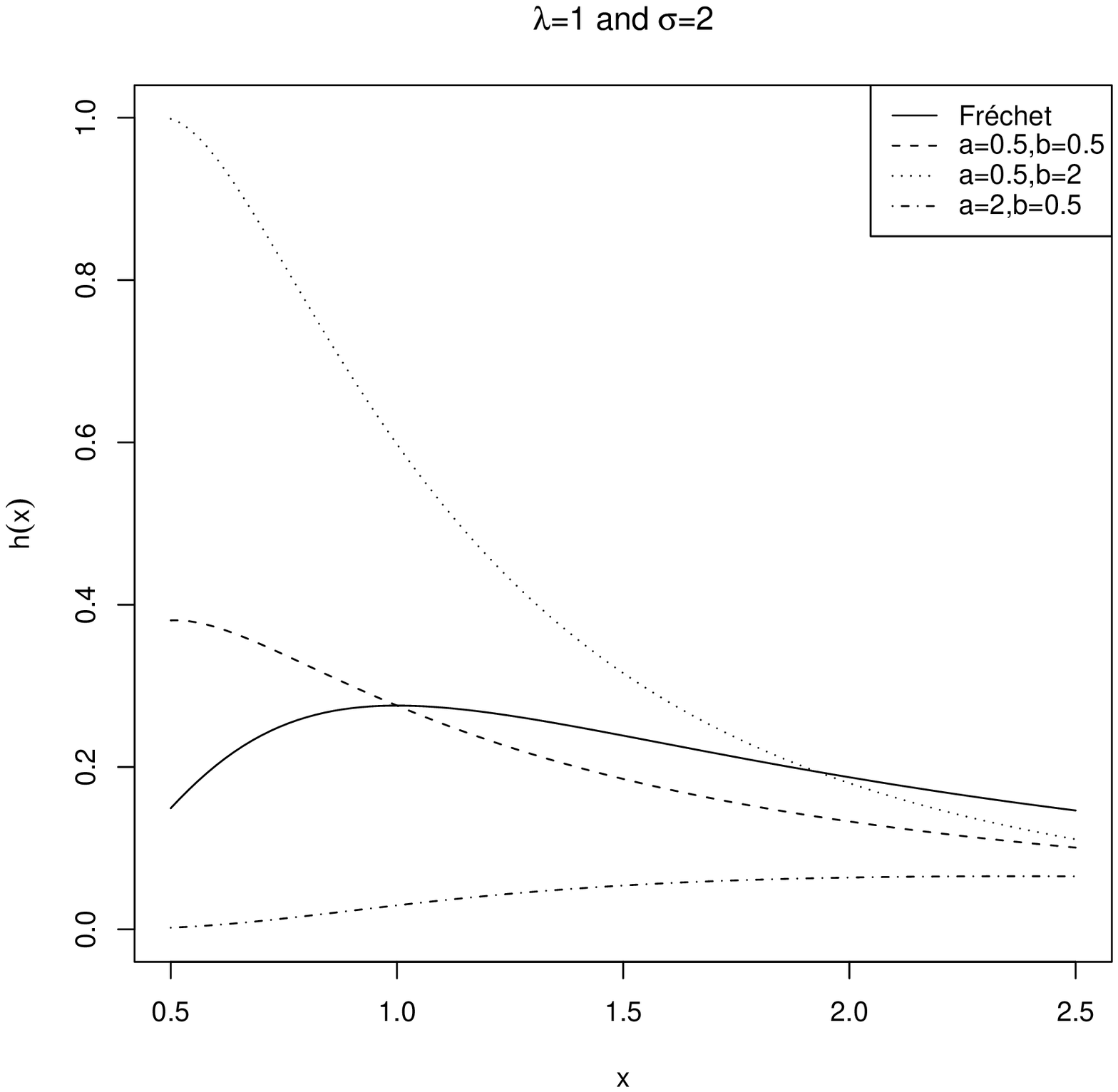}
\includegraphics[width=0.5\textwidth]{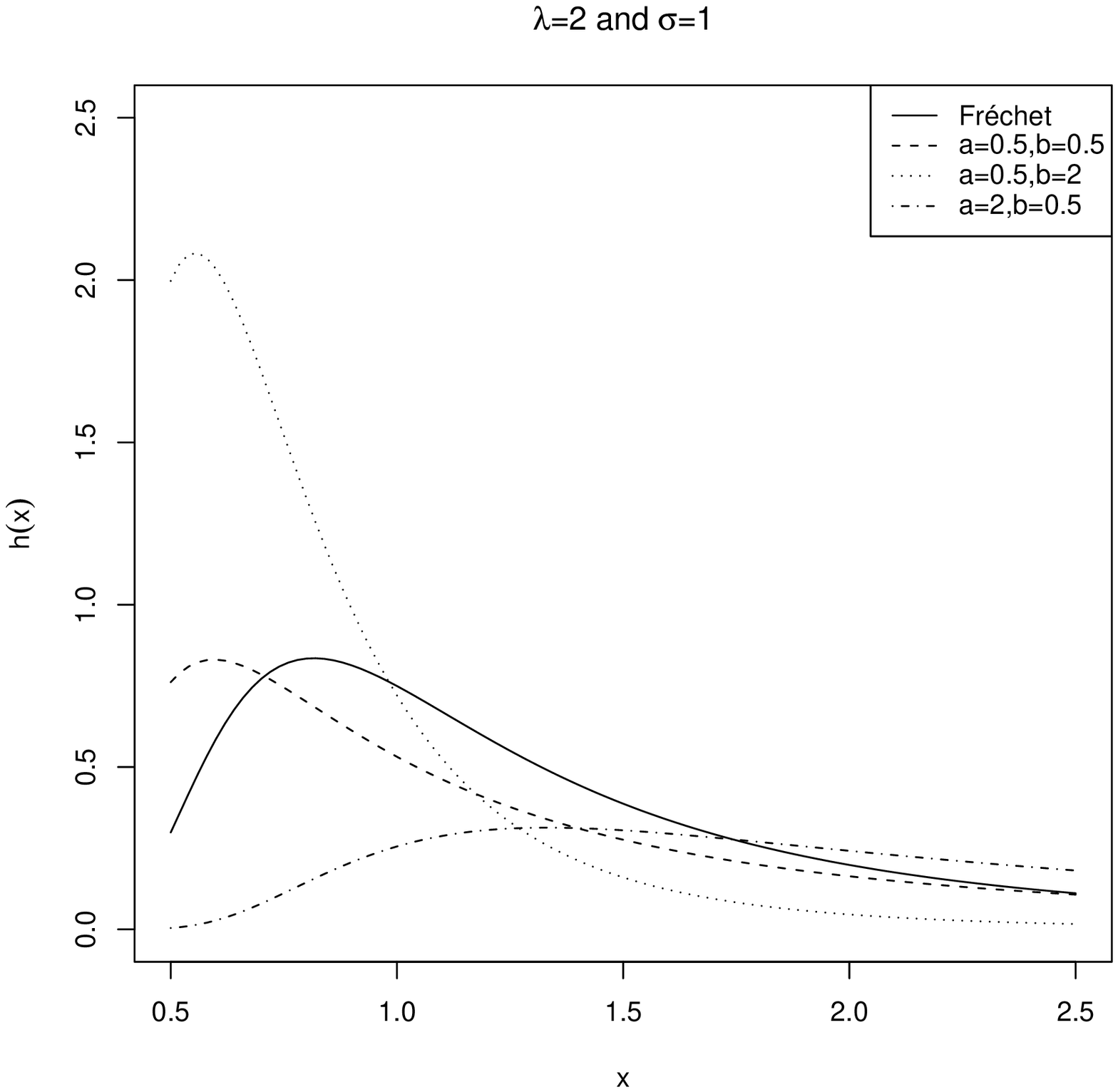}\includegraphics[width=0.5\textwidth]{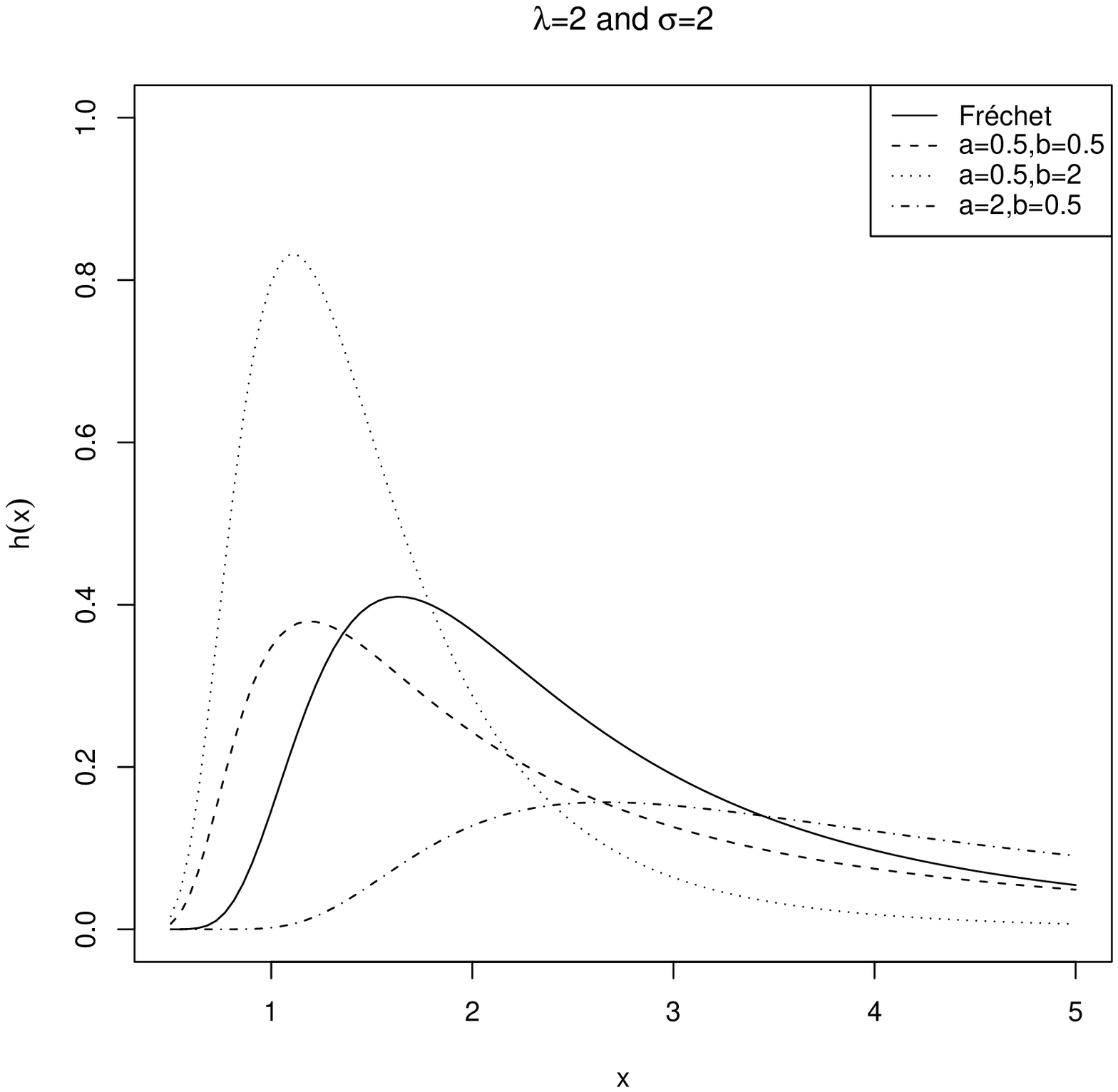}
\caption{Hazard function of the BF distribution for selected parameter values.}
\label{fighaz}
\end{figure}

Nadarajah and Gupta (2004) give the cdf of the BF distribution with
parameters $a>0$, $b>0$, $\sigma>0$ and $\lambda>0$ (denoted by
BF$(a,b,\sigma,\lambda)$) in the same way from (\ref{distfun}) by
replacing the parent cdf $G(x)$ by (\ref{cdffrechet})
\begin{equation}\label{cdfbf}
F(x)=\frac{1}{B(a,b)}\int_0^{\exp\{-(\frac{\sigma}{x})^{\lambda}\}}\omega^{a-1}(1-\omega)^{b-1} d\omega
=I_{\exp\{-(\frac{\sigma}{x})^{\lambda}\}}(a,b),\quad x>0.
\end{equation}
They also give the corresponding pdf and hazard function,
respectively, as
\begin{equation}\label{pdfbf}
f(x)=\frac{\lambda\sigma^{\lambda}}{B(a,b)}x^{-(\lambda+1)}e^{-a(\frac{\sigma}{x})^{\lambda}}\{1-e^{-(\frac{\sigma}{x})^{\lambda}}\}^{b-1},\quad x>0,
\end{equation}
and
\begin{equation}\label{hazbf}
\tau(x)=\frac{\lambda\sigma^{\lambda}x^{-(\lambda+1)}e^{-a(\frac{\sigma}{x})^{\lambda}}\{1-e^{-(\frac{\sigma}{x})^{\lambda}}\}^{b-1}}{B(a,b)\{1-I_{\exp\{-(\frac{\sigma}{x})^\lambda\}}(a,b)\}},\quad x>0.
\end{equation}
Figures \ref{figpdf} and \ref{fighaz} illustrate some of the possible shapes of the
pdf (\ref{pdfbf}) and hazard function (\ref{hazbf}), respectively, for selected
parameter values, including the case of the Fréchet distribution. The BF distribution is
easily simulated from (\ref{cdfbf}) as follows: if $V\sim B(a,b)$ then $X = \sigma/(-{\rm log}V)^{1/\lambda}$
has the BF$(a,b,\sigma,\lambda)$ distribution.

The BF distribution generalizes some well-known distributions. The
exponentiated Fréchet (EF) distribution (Nadarajah and Kotz, 2003)
is a special case when $a=1$. The Fréchet distribution (with
parameters $\sigma$ and $\lambda$) is also a special case of
(\ref{pdfbf}) when $a=b=1$. Further, when $b=1$ and $\lambda=1$,
(\ref{pdfbf}) is an inverse gamma distribution with shape parameter
$2$ and scale parameter $a \sigma$. Since the BF distribution
generalizes the Fréchet and EF distributions by adding two
parameters and one parameter, respectively, it can be used by
practitioners as an extra tool to analyze the data we would normally
use with the last two distributions. The book of Kotz and Nadarajah
(2000) demonstrates the applicability of the Fréchet distribution in
several fields.

The rest of the paper is organized as follows. Section 2 gives
expansions for the pdf and cdf of the BF distribution and for the
density of the order statistics depending on whether the parameters
$b$ (or $a>0$) is real non-integer or integer. We show that the
density functions of the BF and its order statistics can be
expressed as mixture of Fréchet density functions. The moments of
this distribution and of the order statistics are not known and
general expansions are derived in Section 3 for the cases $b>0$ real
non-integer and integer. L-moments (Hosking, 1986) are expectations
of certain linear combinations of order statistics and form the
basis of a general theory which covers the summarization and
description of theoretical probability distributions. In Section 4
we present expansions for the L-moments of the BF distribution. We
discuss in Section 5 maximum likelihood estimation and calculate the
elements of the information matrix. Section 6 provides two
applications to real data sets.
In Section 7 we end with some conclusions.\\

\noindent 2. EXPANSIONS FOR THE DISTRIBUTION AND DENSITY FUNCTIONS\\

Here, we provide simple expansions for the cdf of the BF
distribution depending on whether the parameter $b$ (or $a$) is real
non-integer or integer. We consider the series expansion
\begin{eqnarray}\label{exp}
(1-z)^{b-1}=\sum_{j=0}^\infty
\frac{(-1)^j\Gamma(b)}{\Gamma(b-j)j!}z^j,
\end{eqnarray}
valid for $|z|<1$ and $b>0$ real and non-integer. Application of
(\ref{exp}) to (\ref{distfun}) if $b$ is real non-integer gives
\begin{eqnarray}\label{cdfbreal}
F(x)=\frac{\Gamma(a+b)}{\Gamma(a)}\sum_{j=0}^\infty\frac{(-1)^j
G_{\sigma,\lambda}(x)^{a+j}}{\Gamma(b-j)j!(a+j)},
\end{eqnarray}
where $G_{\sigma,\lambda}(x)$ comes from (\ref{cdffrechet}). Then,
we have
\begin{eqnarray}\label{distbf}
F(x)=\frac{\Gamma(a+b)}{\Gamma(a)}\sum_{j=0}^\infty\frac{(-1)^je^{-(a+j)\left(\frac{\sigma}{x}\right)^\lambda}}{\Gamma(b-j)j!(a+j)}.
\end{eqnarray}
For $b$ integer, the sum in (\ref{distbf}) simply stops at $b-1$.
When $b=1$, it follows $F(x)=e^{-a(\frac{\sigma}{x})^\lambda}$.

It can be seen in the Wolfram Functions Site \footnote{{\tt
http://functions.wolfram.com/}} that for integer $b$
$$I_y(a,b)=\frac{y^a}{\Gamma(a)}\sum_{j=0}^{b-1}\frac{\Gamma(a+j)(1-y)^j}{j!}$$
and for integer $a$
$$I_y(a,b)=1-\frac{(1-y)^b}{\Gamma(b)}\sum_{j=0}^{a-1}\frac{\Gamma(b+j)}{j!}y^j.$$
Hence, if $b$ is integer, we obtain another equivalent form for
(\ref{distbf})
$$F(x)=\frac{e^{-a(\frac{\sigma}{x})^\lambda}}{\Gamma(a)}\sum_{j=0}^{b-1}\frac{\Gamma(a+j)(1-e^{-(\frac{\sigma}{x})^\lambda})^j}{j!}.$$
and, for integer values of $a$, we have
$$F(x)=1-\frac{(1-e^{-(\frac{\sigma}{x})^\lambda})^b}{\Gamma(b)}\sum_{j=0}^{a-1}\frac{\Gamma(b+j)}{j!}e^{-j(\frac{\sigma}{x})^\lambda}.$$
If $a=1$, the above expression reduces to
$$ F(x)=1-\{1-e^{-(\frac{\sigma}{x})^\lambda}\}^b,$$
which agrees with the cdf of the EF distribution.

The pdf in (\ref{pdfbf}) is straightforward to compute using any
statistical software. However, we show that the BF density can be
expressed as an infinite (or finite) weighted linear combination of
pdf's of random variables having Fréchet distributions. This is
important to provide some mathematical properties of the BF
distribution directly from the corresponding properties of the
Fréchet distribution. If $b>0$ is real non-integer, and again using
(\ref{exp}) we can rewrite (\ref{pdfbf}) as
\begin{equation}\label{pdfbfalt}
f(x)=\sum_{k=0}^\infty w_k g_{a_k,\lambda}(x),
\end{equation}
where
$$w_k= \Gamma(a+b)(-1)^k/\{\Gamma(a)\Gamma(b-k) k! (k+a)\}$$
represent weighted constants such that $\sum_{k=0}^\infty w_k=1$ and
$g_{a_k,\lambda}(x)$ is a Fréchet density with scale parameter
$a_k=\sigma(k+a)^{1/\lambda}$ and shape parameter $\lambda$. In
addition, if $a=1$, (\ref{pdfbfalt}) agrees with the corresponding
result obtained by Nadarajah and Kotz (2003, Section 5). If $b>0$ is
integer, the sum in (\ref{pdfbfalt}) is finite and stops at $b-1$.
Then, the ordinary, central, factorial moments and the moment
generating function of the BF distribution could in principle follow
from the same weighted infinite (or finite if $b$ is integer) linear
combination of the corresponding quantities for the Fréchet
distribution.

We now give the density of the $i$th order statistic $X_{i:n}$,
$f_{i:n}(x)$ say, in a random sample of size $n$ from the BF
distribution. It is well known that
\begin{eqnarray*}
f_{i:n}(x)=\frac{1}{B(i,n-i+1)}f(x)F^{i-1}(x)\{1-F(x)\}^{n-i},
\end{eqnarray*}
for $i=1,\ldots,n$. Using (\ref{cdfbf}) and (\ref{pdfbf}) we can express
$f_{i:n}(x)$ in terms of the incomplete beta function ratio
\begin{eqnarray*}
f_{i:n}(x)&=&\frac{n!\,g_{\sigma,\lambda}(x)}{(i-1)!(n-i)!B(a,b)} G_{\sigma,\lambda}(x)^{a-1}
\{1-G_{\sigma,\lambda}(x)\}^{b-1}\times\\
&& {I_{G_{\lambda,\alpha}(x)}(a,b)}^{i-1}\,{I_{\{1-G_{\lambda,\alpha}(x)\}}(b,a)}^{n-i}.
\end{eqnarray*}
The cdf of the $i$th order statistic $X_{i:n}$, $F_{i:n}(x)$ say, is
\begin{eqnarray*}
F_{i:n}(x)&=&\sum_{r=i}^{n}\binom{n}{r} {I_{G_{\lambda,\alpha}(x)}(a,b)}^{r}{I_{\{1-G_{\lambda,\alpha}(x)\}}(b,a)}^{n-r}.
\end{eqnarray*}

Using the identity $(\sum_{k=0}^\infty a_k x^k)^n=\sum_{k=0}^\infty
c_{k,n} x^k$ (see Gradshteyn and Ryzhik, 2000), where
$c_{0,n}=a_0^n$ and
$$c_{k,n}=(k a_0)^{-1}\sum_{l=1}^{k}(n l-k+l)a_l c_{k-l,n}$$ for
$k=1,2,\ldots$ and (\ref{distbf}), the pdf of the $i$th order
statistic can be written for $b>0$ real non-integer and integer,
respectively, as
\begin{eqnarray}\label{exp.gra.real}
f_{i:n}(x)=\sum_{k=0}^{n-i}\sum_{j=0}^{\infty}\frac{(-1)^k\binom{n-i}{k}\Gamma(b)^{i+k-1}B(a(i+k)+j,b)c^{(1)}_{i,j,k}}{B(a,b)^{i+k}B(i,n-i+1)}f_{i,j,k}(x)
\end{eqnarray}
and
\begin{eqnarray}\label{exp.gra.integer}
f_{i:n}(x)=\sum_{k=0}^{n-i}\sum_{j=0}^{b-1}\frac{(-1)^k\binom{n-i}{k}B(a(i+k)+j,b)c^{(2)}_{i,j,k}}{B(a,b)^{i+k}B(i,n-i+1)}f_{i,j,k}(x),
\end{eqnarray}
where $f_{i,j,k}(x)$ is the denstity of a
BF$(a(i+k)+j,b,\sigma,\lambda)$ distribution,
\begin{eqnarray*}
c^{(1)}_{i,0,k}=\left\{\frac{1}{a\Gamma(b)}\right\}^{i+k-1},\quad  c^{(1)}_{i,j,k}=\frac{a\Gamma(b)}{j}\sum_{l=1}^j\frac{(-1)^l\{l(i+k)-j\}}{\Gamma(b-l)l!(a+l)}c_{i,j-l,k}
\end{eqnarray*}
and
\begin{eqnarray*}
c^{(2)}_{i,0,k}=\left(\frac{1}{a}\right)^{i+k-1},\quad
c^{(2)}_{i,j,k}=\frac{a}{j}\sum_{l=1}^j\frac{(-1)^l\binom{b-1}{l}\{l(i+k)-j\}}{a+l}c_{i,j-l,k},
\end{eqnarray*}
for $j\geq1$. Expansions (\ref{distbf})-(\ref{exp.gra.integer}) are
the main results of this section.

Two alternative expansions for the densities of the order statistics
follow from the identity $(\sum_{i=1}^\infty
a_i)^k=\sum_{\{m_1,\ldots,m_k\}=0}^\infty a_{m_1}\ldots a_{m_k}$ for
$k$ a positive integer. Using this identity and (\ref{distbf}), it
is easy to show for $b>0$ real non-integer and integer that
\begin{eqnarray}\label{pdforderreal}
f_{i:n}(x)=\sum_{k=0}^{n-i}\sum_{m_1=0}^{\infty}\ldots\sum_{m_{i+k-1}=0}^{\infty}\delta^{(1)}_{i,k} f_{i,k}(x)
\end{eqnarray}
and
\begin{eqnarray}\label{pdforderinteger}
f_{i:n}(x)=\sum_{k=0}^{n-i}\sum_{m_1=0}^{b-1}\ldots \sum_{m_{i+k-1}=0}^{b-1}\delta^{(2)}_{i,k} f_{i,k}(x),
\end{eqnarray}
respectively, where $f_{i,k}(x)$ is the pdf of a BF$(a(i+k)+\sum_{j=1}^{i+k-1}m_j,b,\sigma,\lambda)$
distribution, $$\delta^{(1)}_{i,k}=\frac{(-1)^{k+\sum_{j=1}^{i+k-1}m_j}\binom{n-i}{k}B(a(i+k)+\sum_{j=1}^{i+k-1}m_j,b)\Gamma(b)^{i+k-1}}{B(a,b)^{i+k}B(i,n-i+1)\prod_{j=1}^{i+k-1}\Gamma(b-m_j)m_j!(a+m_j)}$$ and
$$\delta^{(2)}_{i,k}=\frac{(-1)^{k+\sum_{j=1}^{i+k-1}m_j}\binom{n-i}{k}B(a(i+k)+\sum_{j=1}^{i+k-1}m_j,b)}{B(a,b)^{i+k}B(i,n-i+1)}\prod_{j=1}^{k+j-1}\frac{\binom{b-1}{m_j}}{(a+m_j)}.$$
The summation in (\ref{pdforderreal}) and (\ref{pdforderinteger})
extends over all $(i+k)$-tuples ($k,m_1,\ldots,m_{i+k-1}$) of
non-negative integers and is implementable on a computer. However,
expansions (\ref{exp.gra.real}) and (\ref{exp.gra.integer}) are much
simpler to be calculated and their CPU times are usually
smaller.\\

\noindent 3. MOMENTS\\

As with any other distribution, many of the interesting
characteristics and features of the BF distribution can be studied
through the moments. We obtain immediately the $r$th moment
$\mu_r^{\prime}$ of the BF distribution from (\ref{pdfbfalt}) if
$r<\lambda$
\begin{equation}\label{mominf}
\mu_r^{\prime}=\frac{\sigma^r\Gamma(1-r/\lambda)\Gamma(a+b)}{\Gamma(a)}\sum_{j=0}^\infty\frac{(-1)^j(a+j)^{r/\lambda-1}}{\Gamma(b-j)j!}.
\end{equation}
If $b>0$ is integer and $r<\lambda$, the sum stops at $b-1$. If
$a=1$ and $r<\lambda$, (\ref{mominf}) gives the $r$th moment of the
EF distribution with parameters $b$, $\sigma$ and $\lambda$ which is a new result
for the EF distribution.\\

From (\ref{exp.gra.real}) and (\ref{exp.gra.integer}), we obtain
simple expansions for the moments of the order statistics. The $r$th
moment of the $X_{i:n}$ for $b>0$ real non-integer is
\begin{equation}\label{momentorder1}
E(X^r_{i:n})=\sum_{k=0}^{n-i}\sum_{j=0}^{\infty}\frac{(-1)^k\binom{n-i}{k}\Gamma(b)^{i+k-1}B(a(i+k)+j,b)c^{(1)}_{i,j,k}}{B(a,b)^{i+k}B(i,n-i+1)}E(X^r_{i,j,k})
\end{equation}
and for $b>0$ integer
\begin{equation}\label{momentorder2}
E(X^r_{i:n})=\sum_{k=0}^{n-i}\sum_{j=0}^{b-1}\frac{(-1)^k\binom{n-i}{k}B(a(i+k)+j,b)c^{(2)}_{i,j,k}}{B(a,b)^{i+k}B(i,n-i+1)}E(X^r_{i,j,k}),
\end{equation}
where $X_{i,j,k}\sim$ BF$(a(i+k)+j,b,\sigma,\lambda)$.
From (\ref{pdforderreal}) and (\ref{pdforderinteger}), we obtain two alternative expressions for
the moments of the order statistics. These expressions for $b>0$ real non-integer and integer are given by
\begin{eqnarray*}
E(X_{i:n}^r)=\sum_{k=0}^{n-i}\sum_{m_1=0}^{\infty}\ldots\sum_{m_{i+k-1}=0}^{\infty}\delta^{(1)}_{i,k}E(X_{i,k}^r)
\end{eqnarray*}
and
\begin{eqnarray*}
E(X_{i:n}^r)=\sum_{k=0}^{n-i}\sum_{m_1=0}^{b-1}\ldots\sum_{m_{i+k-1}=0}^{b-1}\delta^{(2)}_{i,k} E(X_{i,k}^r),
\end{eqnarray*}
respectively, where $X_{i,k}\sim BF(a(i+k)+\sum_{j=1}^{i+k-1}m_j,b,\sigma,\lambda)$.\\

Graphical representation of skewness and kurtosis when $\lambda=5$
and $\sigma=1$, as a function of parameter $a$ for some choices of
parameter $b$, and as a function of parameter $b$ for some choices
of parameter $a$, are given in Figures 3 and 4, respectively.\\

\begin{figure}[h!]\label{skew}
\centering
\includegraphics[width=0.5\textwidth]{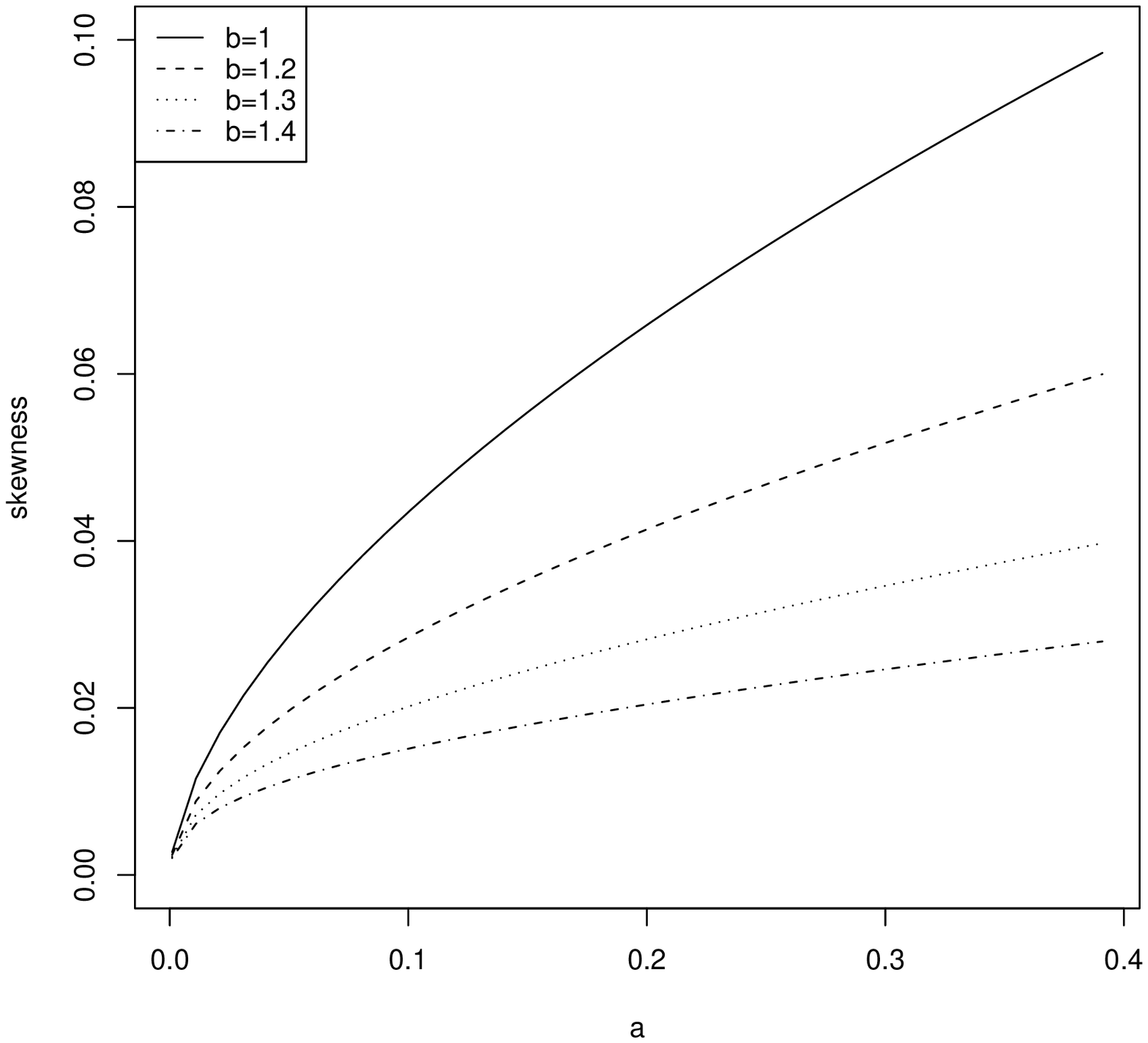}\includegraphics[width=0.5\textwidth]{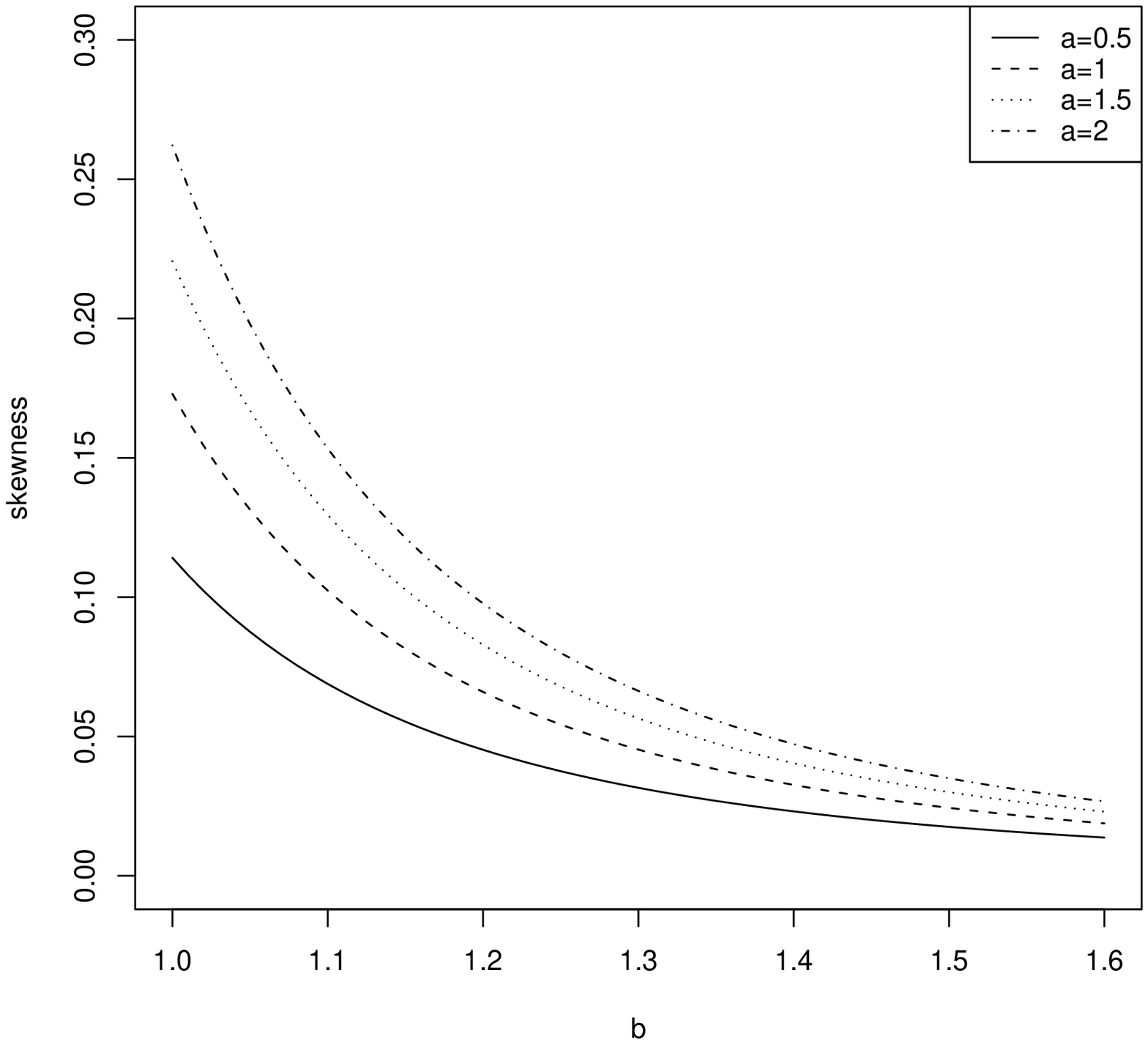}
\caption{Skewness of the BF distribution as a function of $a$($b$) for selected values of $b$($a$).}
\end{figure}

\begin{figure}[h!]\label{kurt}
\centering
\includegraphics[width=0.5\textwidth]{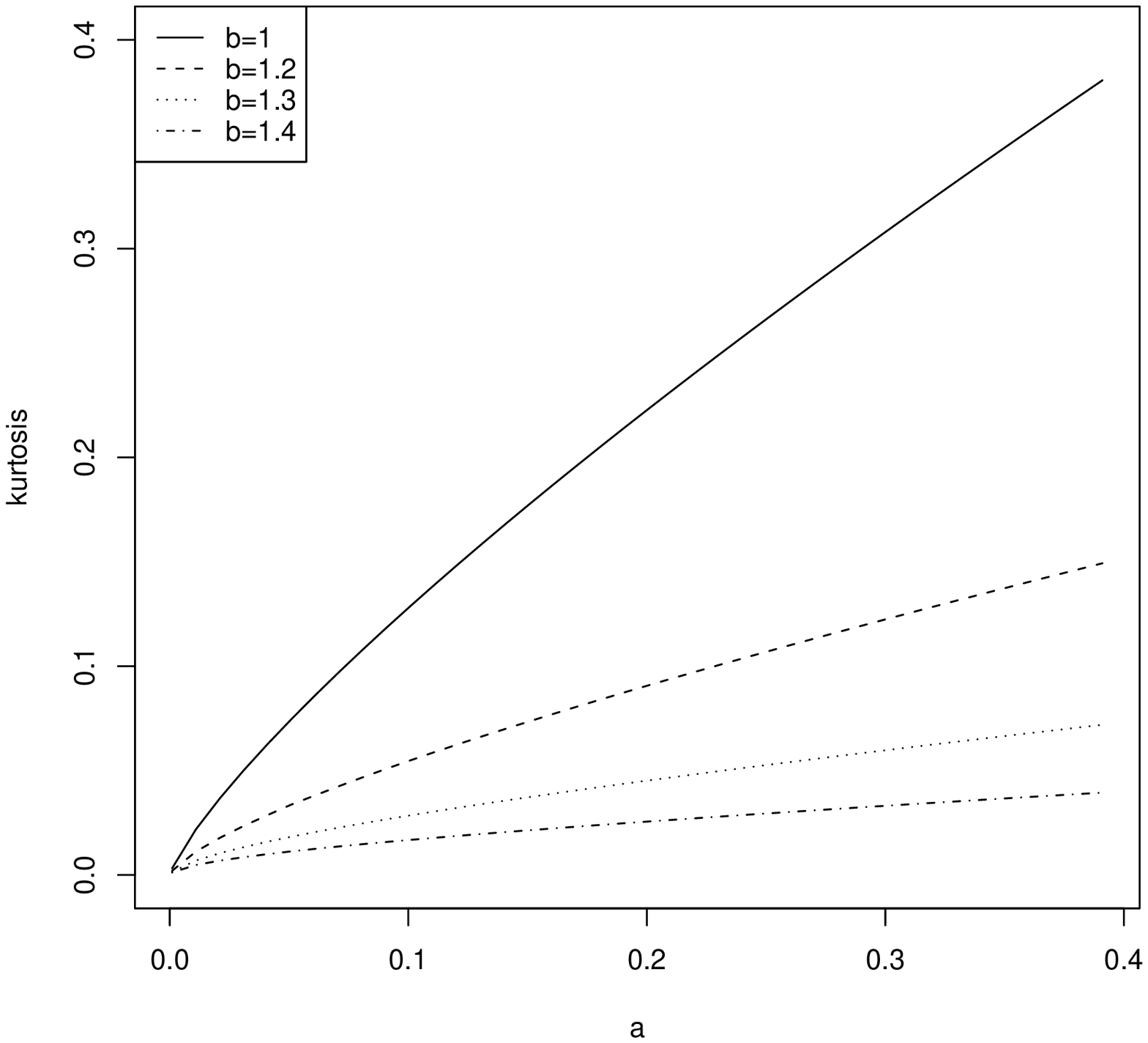}\includegraphics[width=0.5\textwidth]{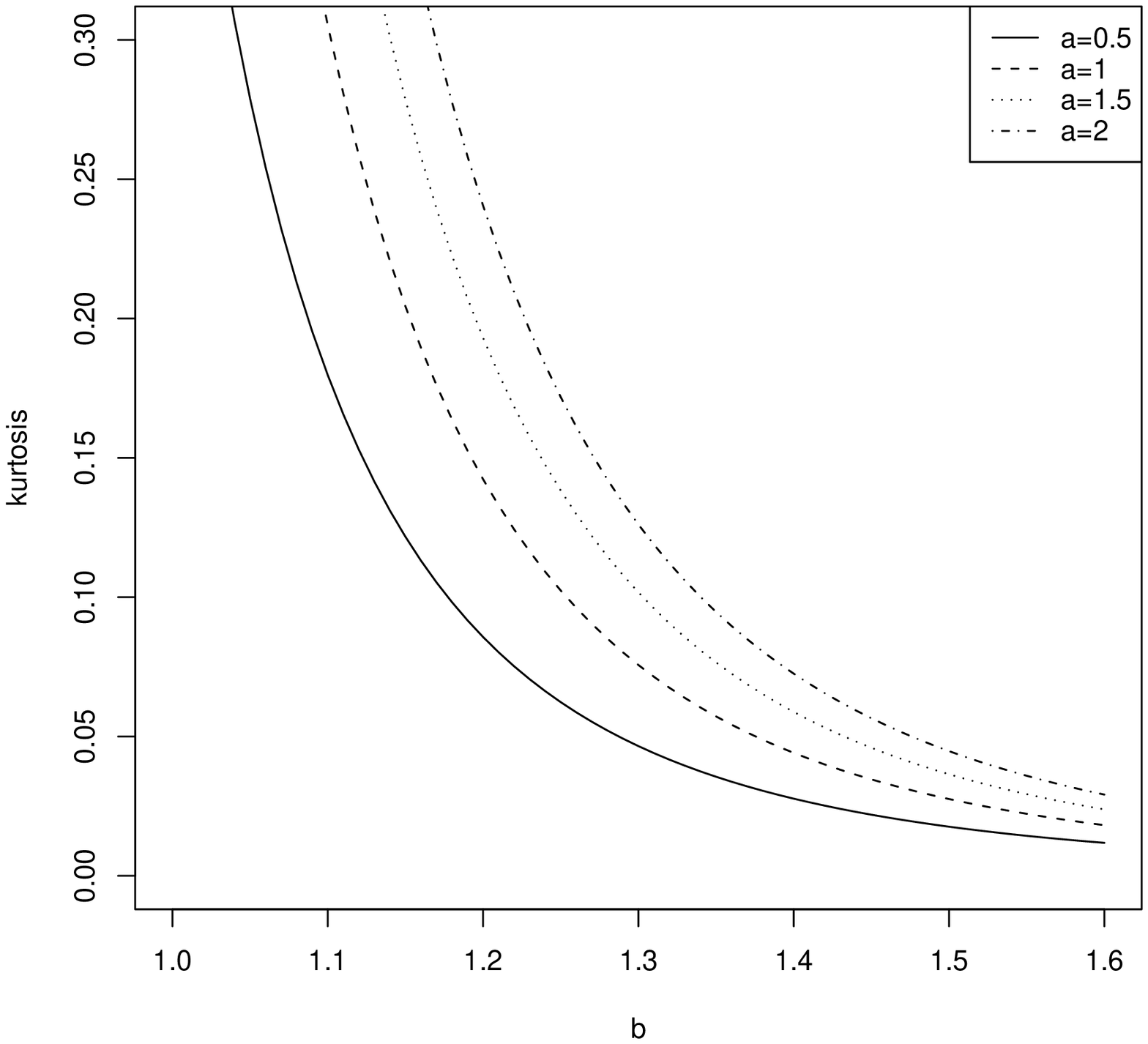}
\caption{Kurtosis of the BF distribution as a function of $a$($b$) for selected values of $b$($a$).}
\end{figure}

\noindent 4. L-MOMENTS\\

The L-moments are analogous to the ordinary moments but can be
estimated by linear combinations of order statistics. They are
linear functions of expected order statistics defined by (Hosking,
1990)
\begin{equation}\label{Lmoments}
\lambda_{r+1} = (r+1)^{-1} \sum_{k=0}^{r} (-1)^k \binom{r}{k}
E(X_{r+1-k:r+1}), \,\,r=0,1,\ldots
\end{equation}
The first four L-moments are: $\lambda_1=E(X_{1:1})$,
$\lambda_2=\frac{1}{2}E(X_{2:2}-X_{1:2})$,
$\lambda_3=\frac{1}{3}E(X_{3:3}-2X_{2:3}+X_{1:3})$ and
$\lambda_4=\frac{1}{4}E(X_{4:4}-3X_{3:4}+ 3X_{2:4}-X_{1:4})$. The
L-moments have the advantage that they exist whenever the mean of
the distribution exists, even though some higher moments may not
exist, and are relatively robust to the effects of outliers.

From the expansions (\ref{momentorder1})-(\ref{momentorder2}) for
the moments of the order statistics we can obtain expansions for the
L-moments of the BF distribution as weighted linear combinations of
the means of suitable BF distributions.\\

\noindent 5. ESTIMATION AND INFORMATION MATRIX\\

We assume that $Y$ follows the BF distribution and let $\theta=(a,b,\sigma,\lambda)^T$
be the true parameter vector. The log-likelihood $\ell=\ell(\theta)$ for a single observation $y$ of $Y$ is given by
$$\ell=\log\lambda+\lambda\log(\sigma/y)-\log\{B(a,b)\}-a (\sigma/y)^\lambda
+(b-1)\log\{1-e^{-(\sigma/y)^\lambda}\}.$$
The components of the score vector
$U=U(\theta)=(\partial \ell/\partial a,\partial\ell/\partial b,\partial\ell/\partial\sigma,\partial\ell/\partial\lambda)^{T}$
for one observation are given by
\begin{eqnarray*}
\frac{\partial \ell}{\partial a}&=&-\psi(a)+\psi(a+b)-(\sigma/y)^\lambda,\\
\frac{\partial \ell}{\partial b}&=&-\psi(b)+\psi(a+b)+\log\{1-e^{-(\frac{\sigma}{y})^\lambda}\},\\
\frac{\partial \ell}{\partial \sigma}&=&\frac{\lambda}{\sigma}-\frac{\lambda\sigma^{\lambda-1}}{y^\lambda}\left\{a-\frac{b-1}{e^{(\frac{\sigma}{y})\lambda}-1}\right\},\\
\frac{\partial \ell}{\partial \lambda}&=&\frac{1}{\lambda}+\log\left(\frac{\sigma}{y}\right)\left[1-\left(\frac{\sigma}{y} \right)^\lambda\left\{a-\frac{b-1}{e^{(\frac{\sigma}{y})\lambda}-1}\right\}\right].
\end{eqnarray*}
From $E(\partial \ell/\partial a)=0$, we obtain
$$
E(X^{-\lambda})=\frac{\psi(a+b)-\psi(a)}{\sigma^\lambda}.
$$
For interval estimation and hypothesis tests on the model parameters, we require the
information matrix. The $4\times 4$ unit information matrix
$K=K(\theta)$ is
$$K = \left(\begin{array}{cccc}
\kappa_{a,a}&\kappa_{a,b}&\kappa_{a,\sigma}&\kappa_{a,\lambda}\\
\kappa_{a,b}&\kappa_{b,b}&\kappa_{b,\sigma}&\kappa_{b,\lambda}\\
\kappa_{a,\sigma}&\kappa_{b,\sigma}&\kappa_{\sigma,\sigma}&\kappa_{\sigma,\lambda}\\
\kappa_{a,\lambda}&\kappa_{b,\lambda}&\kappa_{\sigma,\lambda}&\kappa_{\lambda,\lambda}
\end{array}\right),$$
whose elements are
\begin{eqnarray*}
\kappa_{a,a}&=&\psi^{'}(a)-\psi^{'}(a+b),\quad\kappa_{b,b}=\psi^{'}(b)-\Psi^{'}(a+b),\\
\kappa_{\sigma,\sigma}&=&\frac{\lambda}{\sigma^2}[1+a(\lambda-1)\{\psi(a+b)-\psi(a)\}+(b-1)(\lambda T_{1,1,2,0}-T_{1,1,1,0})],\\
\kappa_{\lambda,\lambda}&=&\frac{1}{\lambda^2}\{1+aT_{0,0,1,2}+(b-1)(T_{1,2,2,2}-T_{1,1,1,2})\},\\
\kappa_{\sigma,\lambda}&=&-\frac{1}{\sigma}[1-a\{\psi(a+b)-\psi(a)+T_{0,0,1,1}\}+(b-1)(T_{1,1,1,0}+T_{1,1,1,1}\\
& &- \lambda T_{1,2,2,0})],\quad \kappa_{a,b}=-\psi^{'}(a+b),\quad\kappa_{a,\lambda}=\frac{1}{\lambda}T_{0,0,1,1},\\
\kappa_{a,\sigma}&=&\frac{\lambda}{\sigma}\{\psi(a+b)-\psi(a)\},\quad
\kappa_{b,\sigma}=-\frac{\lambda}{\sigma}T_{1,1,1,0},\quad\kappa_{b,\lambda}=-\frac{1}{\lambda}T_{1,1,1,1}.
\end{eqnarray*}
Here, we define a random variable $V$ following a $Beta(a,b)$ distribution and
the expected value
\begin{eqnarray*}
T_{i,j,k,l}=E\left[V^{i}(1-V)^{-j}(-\log V)^{k} \{\log(-\log
V)\}^{l}\right],
\end{eqnarray*}
where the integral obtained from the above definition is numerically
determined using MAPLE and MATHEMATICA for any $a$ and $b$. For
example, for $a=1.5$ and $b=2.5$ we easily calculated all $T$'s in
the information matrix: $T_{1,1,2,0}=0.51230070$,
$T_{1,1,1,0}=0.55296103$, $T_{0,0,1,2}=0.62931802$,
$T_{1,2,2,2}=0.43145336$, $T_{1,1,1,2}=0.32124774$,
$T_{0,0,1,1}=0.48641180$, $T_{1,1,1,1}=-0.16152763$ and
$T_{1,2,2,0}=0.86196008$.

For a random sample $y=(y_1,\ldots,y_n)^T$ of size $n$ from $Y$, the
total log-likelihood is
$$\ell_n=\ell_n(\theta)=\sum_{i=1}^n \ell^{(i)},$$ where $\ell^{(i)}$ is
the log-likelihood for the $i$th observation ($i=1,\ldots,n$) as given before.
The total score function is $U_n=U_n(\theta)=\sum_{i=1}^n U^{(i)}$, where $U^{(i)}$
for $i=1,\ldots,n$ has the form given earlier and the total information matrix is
$K_n(\theta)=n K(\theta)$.

The MLE $\hat\theta$ of $\theta$ is numerically determined from the
solution of the nonlinear system of equations $U_n=0$. Under
conditions that are fulfilled for the parameter $\theta$ in the
interior of the parameter space but not on the boundary, the
asymptotic distribution of $\sqrt n
(\hat\theta-\theta)\,\,\,\,\mathrm{is}\,\,\,\,N_4(0,K(\theta)^{-1}).$
The asymptotic multivariate normal $N_4(0,K_n(\hat\theta)^{-1})$
distribution of $\hat\theta$ can be used to cons\-truct approximate
confidence regions for some parameters and for the hazard and
survival functions. In fact, an $100(1-\gamma)\%$ asymptotic
confidence interval for each parameter $\theta_i$ is given by
$$ACI_i=(\hat\theta_i-z_{\gamma/2}\sqrt{\hat\kappa^{\theta_i,\theta_i}},\hat{\theta_i}
+z_{\gamma/2}\sqrt{\hat\kappa^{\theta_i,\theta_i}}),$$
where $\hat\kappa^{\theta_i,\theta_i}$ denotes the $i$th diagonal element of
$K_n(\hat\theta)^{-1}$ for $i=1,2,3,4$ and $z_{\gamma/2}$ is the quantile $1-\gamma/2$
of the standard normal distribution. The asymptotic normality is also useful
for testing goodness of fit of the four parameter BF distribution and for comparing
this distribution with some of its special submodels using the likelihood ratio (LR)
statistic.

We consider the partition $\theta=(\theta_1^T,\theta_2^T)^T$,
where $\theta_1$ is a subset of the parameters of interest of the BF and
$\theta_2$ is a subset of the remaining parameters. The LR statistic
for testing the null hypothesis $H_0:\theta_1 =\theta_1^{(0)}$
versus the alternative hypothesis $H_1:\theta_1 \neq \theta_1^{(0)}$ is given
by $w= 2\{\ell(\hat{\theta})-\ell(\tilde{\theta})\}$, where $\tilde\theta$ and
$\hat\theta$ denote the MLEs under the null and alternative hypotheses,
respectively. The statistic $w$ is asymptotically (as $n\to\infty$)
distributed as $\chi_k^2$, where $k$ is the dimension of the subset $\theta_1$
of interest. Then, we can compare for example a BF model against an EF
model by testing $H_0:a=1$ versus $H_1:a \ne 1$. We can also compare a $BF$ model
against the Fréchet model by testing $H_0:a=b=1$ versus
$H_1:{\rm \,\,H_0\,\,is\,\,false}$.\\

\noindent 6. APPLICATIONS\\

In this section we fit the BF distribution to two examples of real data and test two
types of hypotheses: $H_0:\mbox{Fréchet}\,\times\,H_1:BF$ and
$H_0:EF\,\times\,H_1:BF$. The first example is an uncensored data set from
Nichols and Padgett (2006) consisting of 100 observations on breaking stress of carbon
fibres (in Gba): 3.7,
2.74, 2.73, 2.5, 3.6, 3.11, 3.27, 2.87, 1.47, 3.11, 4.42, 2.41,
3.19, 3.22, 1.69, 3.28, 3.09, 1.87, 3.15, 4.9, 3.75, 2.43, 2.95,
2.97, 3.39, 2.96, 2.53, 2.67, 2.93, 3.22, 3.39, 2.81, 4.2,  3.33,
2.55, 3.31, 3.31, 2.85, 2.56, 3.56, 3.15, 2.35, 2.55, 2.59, 2.38,
2.81, 2.77, 2.17, 2.83, 1.92, 1.41, 3.68, 2.97, 1.36, 0.98, 2.76,
4.91, 3.68, 1.84, 1.59, 3.19, 1.57, 0.81, 5.56, 1.73, 1.59, 2, 1.22,
1.12, 1.71, 2.17, 1.17, 5.08, 2.48, 1.18, 3.51, 2.17, 1.69, 1.25,
4.38, 1.84, 0.39, 3.68, 2.48, 0.85, 1.61, 2.79, 4.7,  2.03, 1.8,
1.57, 1.08, 2.03, 1.61, 2.12,
1.89, 2.88, 2.82, 2.05, 3.65.\\

The MLEs and the maximized log-likelihood using the BF distribution
are
\begin{eqnarray*}
\hat{a}=0.4108,\quad \hat{b}= 125.1891,\quad
\hat{\lambda}=0.7496,\quad\hat{\sigma}= 31.4556,\quad
\hat{\ell}_{BF}=-142.9640,
\end{eqnarray*}
whereas for the EF and Fréchet distributions we obtain
\begin{eqnarray*}
\quad \hat{b}=52.0491 ,\quad
\hat{\lambda}=0.6181,\quad\hat{\sigma}=26.1730,\quad
\hat{\ell}_{EF}= -145.0870,
\end{eqnarray*}
and
\begin{eqnarray*}
\quad
\hat{\lambda}=1.7690,\quad\hat{\sigma}=1.8916,\quad\hat{\ell}_{\mbox{Fréchet}}=-173.1440,
\end{eqnarray*}
respectively.\\

The second data set is obtained from Smith and Naylor (1987). The data are the strengths of 1.5 cm glass fibres, measured at the National Physical Laboratory, England. Unfortunately, the units of measurement are not given in the paper. The data set is: 
0.55, 0.93, 1.25, 1.36, 1.49, 1.52, 1.58, 1.61, 1.64, 1.68, 1.73, 1.81, 2 ,0.74, 1.04, 1.27, 1.39, 1.49, 1.53, 1.59, 1.61, 1.66, 1.68, 1.76, 1.82, 2.01, 0.77, 1.11, 1.28, 1.42, 1.5, 1.54, 1.6, 1.62, 1.66, 1.69, 1.76, 1.84, 2.24, 0.81, 1.13, 1.29, 1.48, 1.5, 1.55, 1.61, 1.62, 1.66, 1.7, 1.77, 1.84, 0.84, 1.24, 1.3, 1.48, 1.51, 1.55, 1.61, 1.63, 1.67, 1.7, 1.78, 1.89.\\

Fitting the BF, EF and Fréchet distributions we obtain the MLEs and the maximized
log-likelihood:
\begin{eqnarray*}
 \hat{a}=0.3962,\quad \hat{b}=225.7272,\quad \hat{\lambda}=6.8631,\quad\hat{\sigma}=1.3021,\quad\hat{\ell}_{BF}=-90.5180,
\end{eqnarray*}
\begin{eqnarray*}
\hat{b}=112.5986,\quad \hat{\lambda}=7.7859,\quad
\hat{\sigma}=0.9814,\quad\hat{\ell}_{EF}=-93.1962
\end{eqnarray*}
and
\begin{eqnarray*}
 \hat{\lambda}=1.2643,\quad \hat{\sigma}=2.8875,\quad \hat{\ell}_{\mbox{Fréchet}}=-117.7765,
\end{eqnarray*}
respectively.\\

For the first data set, the values of the LR statistics for testing the hypotheses $H_0:\mbox{Fréchet}\,\times\,H_1:BF$
and $H_0:EF\,\times\,H_1:BF$ are: 60.36 (p-value=$7.81\times10^{-14}$) and 4.246 (p-value=$3.93\times10^{-2}$),
respectively. For the second data set, we obtain the values of the LR statistics 54.5170 (p-value=$1.45\times10^{-12}$) and 5.3564 (p-value=$2.06\times10^{-2}$) for the hypotheses $H_0:\mbox{Fréchet}\,\times\, H_1:BF$ and
$H_0:EF \,\times\, H_1:BF$, respectively. Therefore, in both situations, using any usual significance level we reject the null hypotheses in favor of the alternative hypothesis that the BF distribuiton is an
adequate model.

\begin{figure}[h!]
\centering
\includegraphics[width=0.50\textwidth]{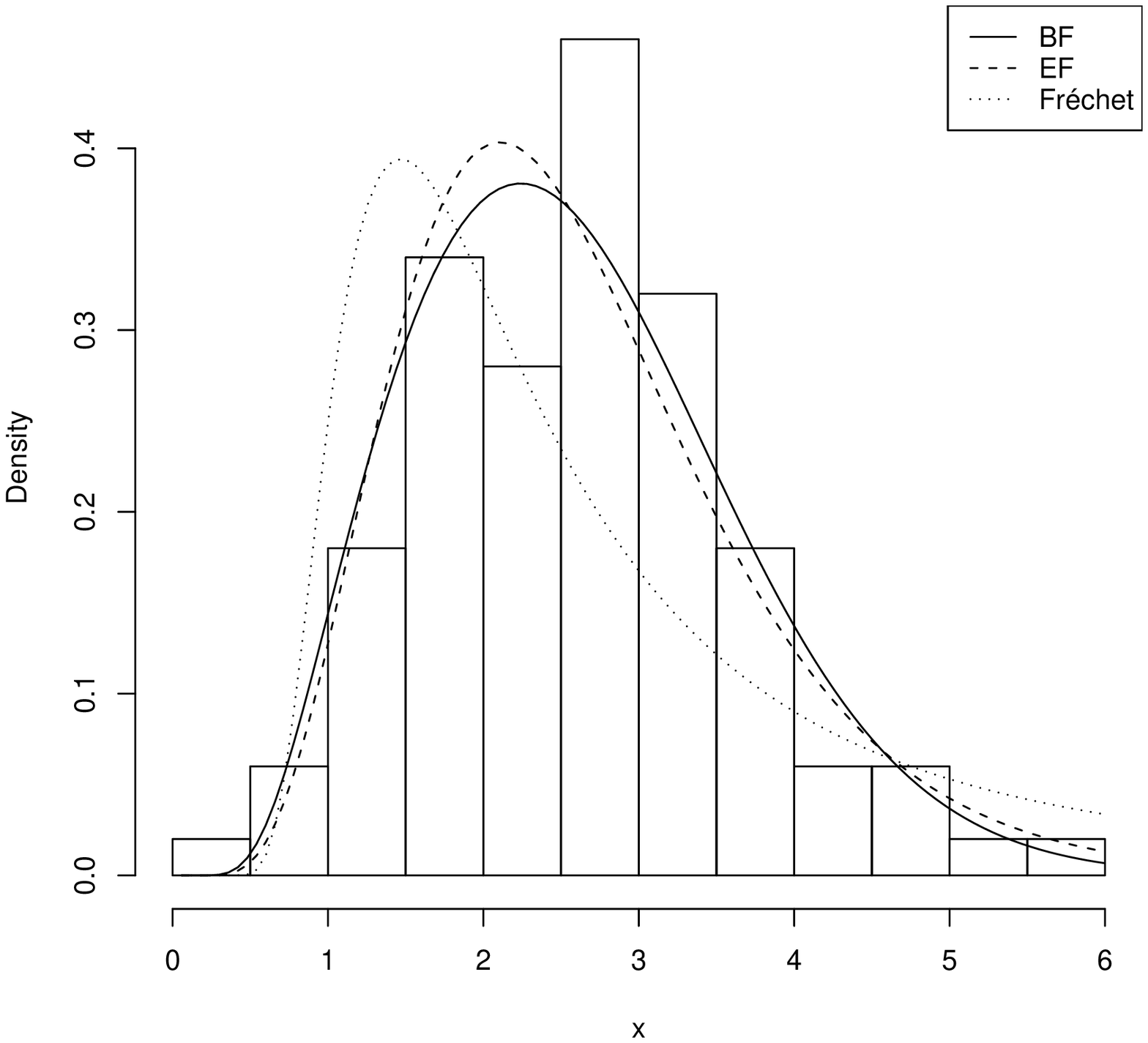}\includegraphics[width=0.50\textwidth]{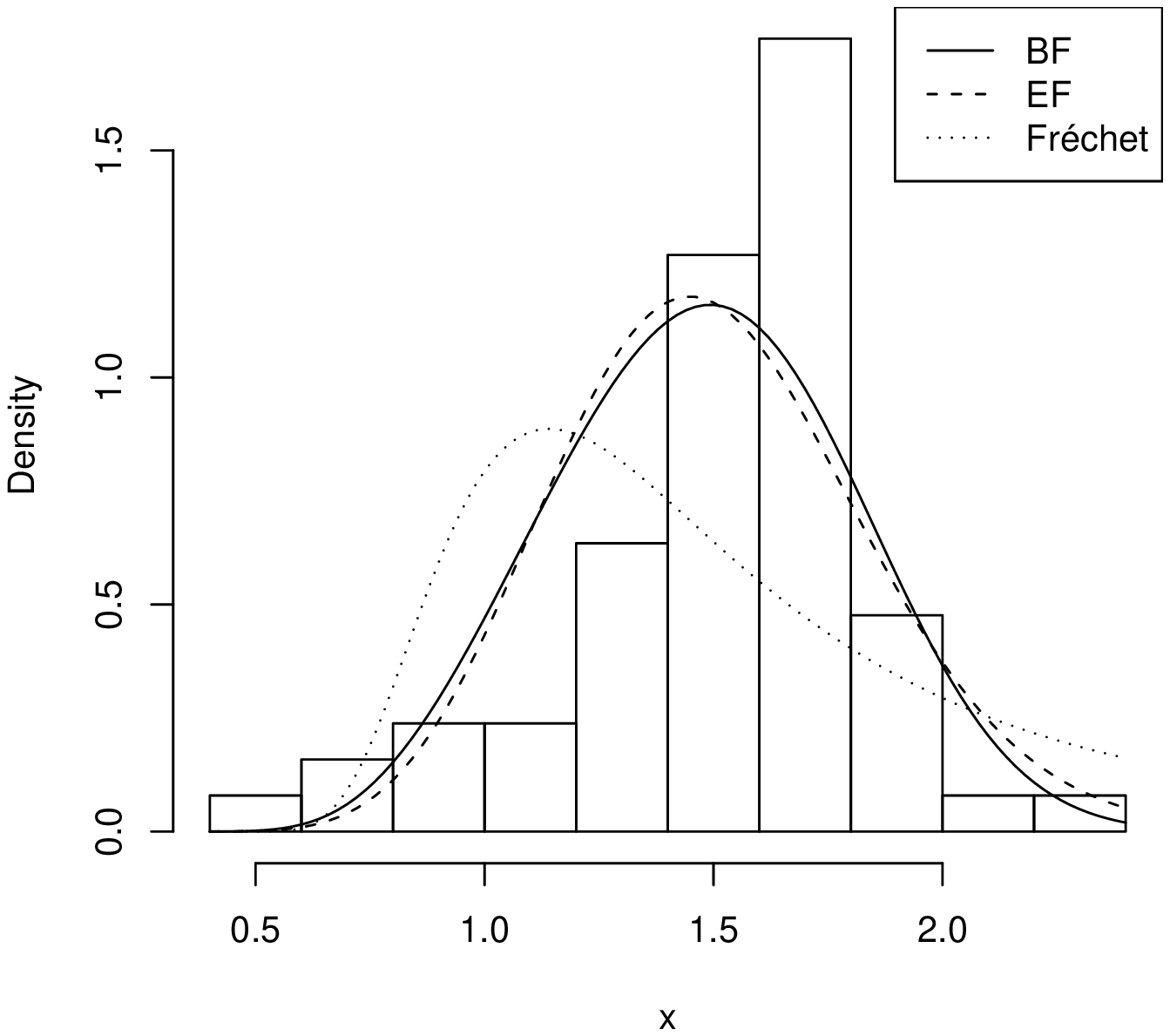}
\caption{Fitted densities of the BF, EF and Fréchet distributions for the data sets 1 and 2, respectively.}
\label{fig-nichols}
\end{figure}

The plots of the estimated densities of the BF, EF and Fréchet distributions given in
Figure \ref{fig-nichols} show that the BF distribution gives a
better fit than the other two sub\-mo\-dels for both data sets.\\

\noindent 7. CONCLUSIONS\\

The BF distribution provides a rather general and flexible framework
for statistical ana\-ly\-sis of positive data. It unifies some
previously proposed distributions, therefore yielding a general
overview of these distributions for theoretical studies, and it also
provides a rather flexible mechanism for fitting a wide spectrum of
real world data sets. The BF distribution is motivated by the wide
use of the Fréchet distribution in practice, and also for the fact
that the generalization provides more flexibility to analyze more
complex situations. In fact, the BF distribution (\ref{pdfbf})
represents a generalization of some distributions previously
considered in the literature such as the Fréchet and EF (Nadarajah
and Kotz, 2003) distributions. This generalization provides a
continuous crossover towards cases with different shapes (e.g.
skewness and kurtosis).

The BF density can be expressed in the mixture form of Fréchet
densities. For doing this, we derived some expansions for the cdf of
the BF distribution and their ordinary and L-moments. We call
attention for the fact that the moments of the EF are not known in
the literature and we derived these moments as a particular case of
our results. The pdf of the BF order statistics can also be
expressed in terms of a linear combination of Fréchet densities. We
also derive the moments of the order statistics. We discuss the
maximum likelihood estimation and obtain the information matrix, and
considered the LR test which may be very useful in practice. We show
that the formulae related with the BF are manageable, and with the
use of modern computer resources with analytic and numerical
capabilities, may turn into adequate tools comprising the arsenal of
applied statisticians. Two numerical examples illustrate that the BF
distribution provides better fits than
the EF and Fréchet distributions.\\

\vskip 5mm

\noindent BIBLIOGRAPHY
\vskip 3mm

\noindent Birnbaum, Z.W., Saunders, S.C. (1969). Estimation for a family of life distribution with applications to fatigue. Journal of Applied Probability 6:328-347.\\

\noindent Eugene, N., Lee, C., Famoye, F. (2002). Beta-normal distribution and its applications. Commun. Statist. - Theory and Methods 31:497-512.\\

\noindent Gradshteyn, I.S., Ryzhik, I.M. (2000). Table of integrals,
series, and pro\-ducts. Academic Press, San Diego.\\

\noindent Gupta, A.K., Nadarajah, S. (2004). On the moments of the beta normal distribution. Commun. Statist. - Theory and Methods 33:1-13.\\

\noindent Hosking, J.R.M. (1990). L-moments: analysis and estimation of distributions using linear
combinations of order statistics. J. Royal Statist. Soc. B 52:105-124.\\

\noindent Jones, M.C. (2004). Families of distributions arising from distributions of order statistics. Test 13:1-43.\\

\noindent Kotz, S., Nadarajah, S. (2000). Extreme Value Distributions: Theory and Applications. Imperial College Press.\\

\noindent Nadarajah, S., Kotz, S. (2003). The exponentiated Fréchet distribution. InterStat. Available online at {\it http://interstat.statjournals.net/YEAR/2003/abstracts/0312001.php}.\\

\noindent Nadarajah, S., Gupta, A.K. (2004). The beta Fréchet distribution. Far East Journal of Theoretical Statistics 14:15-24.\\

\noindent Nadarajah, S., Kotz, S. (2004). The beta Gumbel distribution. Math. Probab. Eng. 10:323-332.\\

\noindent Nadarajah, S. and Kotz, S. (2005). The beta exponential distribution. Reliability Engineering and System Safety 91:689-697.\\

\noindent Smith, R. L. and Naylor, J.C. 1987. A comparison of maximum likelihood and Bayesian estimators for the three-parameter Weibull distribution. Applied Statistics 36: 358-369.

\end{document}